\documentclass[aps,prb,twocolumn,superscriptaddress]{revtex4-2}
\usepackage{graphicx}
\usepackage{mathrsfs}
\usepackage{bm}
\usepackage{amsmath}
\usepackage{color}
\usepackage{dcolumn}
\usepackage{epstopdf}
\usepackage{dsfont}
\usepackage{amssymb}
\usepackage{tabularx}
\usepackage{array,ulem}
\usepackage{float}
\usepackage{colordvi}
\usepackage{verbatim}
\usepackage{hyperref}
\hypersetup{
	colorlinks=true,
	linkcolor=blue,
	filecolor=blue,
	urlcolor=blue,
	citecolor=blue,
}

\begin{document}
\title{
Second-order topological insulator induced by compensated altermagnetism without bulk spin splitting
}

\author{Lizhou Liu}
\affiliation{College of Physics, Hebei Normal University, Shijiazhuang 050024, China}
\affiliation{International Center for Quantum Materials, School of Physics, Peking University, Beijing 100871, China}

\author{Qing-Feng Sun}
\affiliation{International Center for Quantum Materials, School of Physics, Peking University, Beijing 100871, China}
\affiliation{Hefei National Laboratory, Hefei 230088, China}

\author{Ying-Tao Zhang}
\email[Correspondence author:~~]{zhangyt@mail.hebtu.edu.cn}
\affiliation{College of Physics, Hebei Normal University, Shijiazhuang 050024, China}

\date{\today}

\begin{abstract}
We theoretically demonstrate a second-order topological insulating phase induced by compensated altermagnetism, while keeping the bulk gap unchanged, in a two-dimensional topological insulator film. By introducing a layer-resolved out-of-plane $d$-wave altermagnetic term with opposite signs on the top and bottom layers, the system preserves $\mathcal{PT}$ symmetry and maintains spin degeneracy in the bulk bands, while simultaneously gapping the helical edge states and generating localized corner states. The resulting higher-order phase is characterized by nonzero mirror-graded winding numbers, and an effective edge theory shows that the corner states arise from Dirac mass domain walls. We further determine the phase boundaries analytically and construct the corresponding topological phase diagram, establishing a robust route to higher-order topology without bulk spin splitting.
\end{abstract}

\maketitle
\section{Introduction}

Topological insulators are characterized by insulating bulk states and gapless boundary excitations protected by nontrivial bulk band topology~\cite{Haldane1988, Kane2005, Kane2005a, Bernevig2006, Qi2011, Hasan2010, Ren2016}. 
This bulk-boundary correspondence has been extended to higher-order topological phases, in which an $n$-dimensional bulk hosts $(n-d)$-dimensional boundary states with $d\ge 2$, such as corner states in two dimensions and hinge states in three dimensions~\cite{Benalcazar2017, Benalcazar2017a, Li2020, Benalcazar2019, Schindler2018, Langbehn2017, Schindler2018a, Peterson2018, Xu2019, Chen2020a, Liu2024, Liu2024b, Liu2019, Park2019}. 
In two-dimensional second-order topological insulators, both the bulk and the one-dimensional edge states are gapped, while zero-dimensional corner states emerge within the boundary gap~\cite{Lee2020, Sheng2019, Liu2024a, Miao2023}. 
These higher-order topological phases have attracted broad interest in electronic, photonic, and phononic systems because they provide a new route to robust boundary localization and controllable topological functionalities~\cite{Xie2019, Zhu2021, Zhu2022, Huang2023, Han2024}.

A common route to second-order topology is to gap the helical edge states of a first-order topological insulator by introducing symmetry-allowed Dirac mass terms, which in most existing schemes originate from in-plane magnetism~\cite{Ren2020, Miao2024, Chen2020} or related symmetry-breaking perturbations~\cite{Li2024, Miao2025, Liu2025}. 
However, selectively magnetizing only the boundary is experimentally challenging, whereas a more uniform in-plane magnetic perturbation typically affects both edge and bulk states simultaneously~\cite{Ren2020}. 
As a result, such perturbations tend to lift the bulk-band double degeneracy and may even reduce or close the bulk gap, leading to a competition between opening an edge gap and preserving the bulk gap. 
An important open question is therefore whether higher-order topology can be induced without changing the bulk gap.

Altermagnets constitute a distinct class of collinear compensated magnets, fundamentally different from conventional ferromagnets and antiferromagnets~\cite{Smejkal2022, Smejkal2022a, Ahn2019, Smejkal2022b, Yuan2020, Smejkal2020}. 
They typically exhibit momentum-dependent spin splitting while carrying zero net magnetization and have recently attracted broad interest in magnetic, topological, and transport phenomena~\cite{Sun2023, Zhou2023, Zhang2023, Betancourt2023, Yi2026, Sun2025a, Wan2025, Liu2025t, Cheng2024, Liu2026q, Sun2025t}. 
Recent works further suggest that layered systems provide a natural platform for engineering altermagnetic order~\cite{Zeng2024, Pan2024, Liu2024c}. 
In particular, compensated altermagnetism has been proposed in layered structures, where finite local altermagnetic spin splitting is canceled globally by an additional symmetry relation between different layers, such that the bulk bands remain doubly degenerate without bulk spin splitting~\cite{Meier2026}. 
This property makes it fundamentally different from conventional magnetic mechanisms, which typically gap boundary modes while simultaneously lifting the bulk double degeneracy. 
It therefore suggests a promising route to higher-order topology by opening an edge gap while preserving an unsplit bulk spectrum.

In this work, we study a topological insulator film subject to a compensated $d$-wave altermagnetic term with opposite signs in the top and bottom layers. Although this term preserves $\mathcal{PT}$ symmetry and therefore does not induce bulk spin splitting, it gaps the helical edge states of the parent topological insulator film and drives the system into a two-dimensional second-order topological insulating phase hosting zero-energy corner states. We show that this higher-order phase is characterized by nonzero mirror-graded winding numbers defined along the diagonal high-symmetry lines, and that the corner states originate from Dirac mass domain walls between adjacent edges. We further determine the phase boundaries analytically and construct the corresponding topological phase diagram. Our results establish compensated altermagnetism as an effective route to engineering second-order topology in a $\mathcal{PT}$-symmetric system without bulk spin splitting.

\section{System Hamiltonian}

\begin{figure}[tbp]
  \centering
  \includegraphics[width=8.5cm,angle=0]{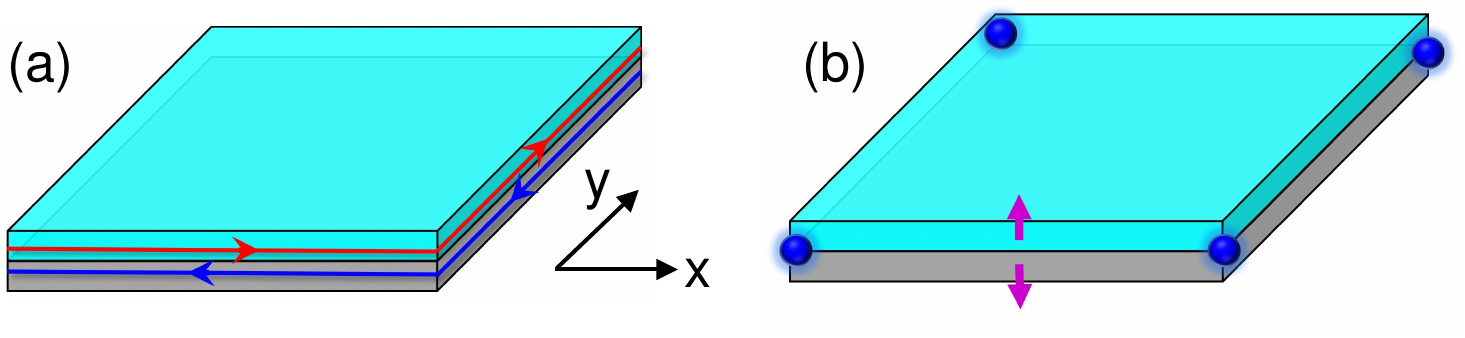}
\caption{(a) Schematic illustration of a topological-insulator film in the quantum spin Hall regime, hosting helical edge states. Cyan and grey denote the top and bottom layers, respectively. (b) Schematic illustration of a second-order topological insulating phase induced by compensated out-of-plane $d$-wave altermagnetism. Blue spheres indicate corner states, and the purple arrows denote the opposite altermagnetic orientations. }
  \label{fig1}
\end{figure}

We consider a Bi$_2$Se$_3$ topological insulator film described by the low-energy effective Hamiltonian for the coupled top and bottom surface states~\cite{Wang2014, Lu2010}. The parent system is chosen in the quantum spin Hall regime and supports helical edge states, as illustrated schematically in Fig.~\ref{fig1}(a).  To investigate the effect of compensated altermagnetism, we introduce a layer-resolved $d$-wave altermagnetic term with opposite signs on the top and bottom layers. This layer-opposite configuration preserves $\mathcal{PT}$ symmetry, such that the bulk bands remain spin degenerate, while still gapping the helical edge states and driving the system into a second-order topological phase, as shown in Fig.~\ref{fig1}(b).

In real space, the tight-binding Hamiltonian of the topological-insulator film is written as~\cite{Shafiei2024, Li2020a}
\begin{align}
H_{\rm TI}
=& \sum_\mathbf{x}\Bigl[\psi_\mathbf{x}^\dagger T_0 \psi_\mathbf{x}
+\bigl(\psi_\mathbf{x}^\dagger T_x \psi_{\mathbf{x}+\delta \hat{x}}
+\psi_\mathbf{x}^\dagger T_y \psi_{\mathbf{x}+\delta \hat{y}}
+ {\rm H.c.}\bigr)\Bigr],
\nonumber\\
T_0
=& (m_0+4m_1)s_0\tau_x,
\nonumber\\
T_x
=& \frac{\upsilon_F}{2i}s_y\tau_z-m_1 s_0\tau_x,
\nonumber\\
T_y
=& -\frac{\upsilon_F}{2i}s_x\tau_z-m_1 s_0\tau_x,
\label{eq:HTI_real}
\end{align}
where $\mathbf{x}$ denotes a lattice site, and $\delta \hat{x}$ and $\delta \hat{y}$ are unit vectors along the $x$ and $y$ directions, respectively. The four-component operator $\psi_\mathbf{x}$ is defined in the basis $\{|T_\uparrow\rangle, |T_\downarrow\rangle, |B_\uparrow\rangle, |B_\downarrow\rangle\}$, where $T$ and $B$ label the top and bottom layers. The Pauli matrices $s_i$ and $\tau_i$ act on the spin and layer degrees of freedom, respectively. Here $m_0$ characterizes the intersurface hybridization, $m_1$ controls the quadratic dispersion term, and $\upsilon_F$ denotes the Fermi velocity.
After Fourier transformation, Eq.~(\ref{eq:HTI_real}) becomes
\begin{align}
H_{\rm TI}(\mathbf{k})
=& \upsilon_F(\sin k_y\, s_x\tau_z-\sin k_x\, s_y\tau_z)
\nonumber\\
&+\Bigl[m_0+2m_1(2-\cos k_x-\cos k_y)\Bigr]\tau_x .
\label{eq:HTI_k}
\end{align}
For the parameter regime considered below, this parent Hamiltonian lies in the quantum spin Hall phase. 
We then introduce a compensated out-of-plane $d$-wave altermagnetic term
\begin{align}
H_{\rm AM}(\mathbf{k})
= J(\cos k_x-\cos k_y)s_z\tau_z,
\label{eq:HAM_k}
\end{align}
where $J$ denotes the altermagnetic strength. The form factor $\cos k_x-\cos k_y$ represents the characteristic $d$-wave anisotropy~\cite{Jiang2025, Ahn2019}, while $s_z$ indicates an out-of-plane spin polarization. The factor $\tau_z$ implies that this term takes opposite signs on the top and bottom layers, corresponding to a compensated layer-opposite altermagnetic configuration~\cite{Meier2026}. 

The total Hamiltonian is therefore
\begin{align}
H(\mathbf{k})
=& \upsilon_F(\sin k_y\, s_x\tau_z-\sin k_x\, s_y\tau_z)
\nonumber\\
&+\Bigl[m_0+2m_1(2-\cos k_x-\cos k_y)\Bigr]\tau_x
\nonumber\\
&+J(\cos k_x-\cos k_y)s_z\tau_z .
\label{eq:Hk_total}
\end{align}
This Hamiltonian preserves $\mathcal{PT}$ symmetry. The inversion operator is $\mathcal{P}=\tau_x$, which exchanges the two layers, and the time-reversal operator is $\mathcal{T}=is_y\mathcal{K}$, where $\mathcal{K}$ denotes complex conjugation. One then finds
\begin{align}
(\mathcal{PT})H(\mathbf{k})(\mathcal{PT})^{-1}=H(\mathbf{k}),
\qquad
(\mathcal{PT})^2=-1.
\label{eq:PT}
\end{align}
As a consequence, the bulk bands remain doubly degenerate at each $\mathbf{k}$ point and exhibit no bulk spin splitting, consistent with the bulk spectra shown in Fig.~\ref{fig2}(a).
For the parameter regime considered here, the bulk band inversion occurs at the $\Gamma$ point. Expanding the Hamiltonian around $\Gamma$, the compensated altermagnetic term takes the form $-(J/2)(k_x^2-k_y^2)s_z\tau_z$, which vanishes exactly at $\Gamma$. Therefore, the bulk gap at the band-inversion point is unchanged by $J$. Correspondingly, a finite $J$ does not modify the bulk gap-closing condition.
In the following, we demonstrate that although the compensated altermagnetic term does not lift the bulk spin degeneracy, it gaps the helical edge states and induces corner states in finite samples.


\section{Results and Discussion}

\subsection{Second-order topological insulator}

\begin{figure}
  \centering
  \includegraphics[width=8.5cm,angle=0]{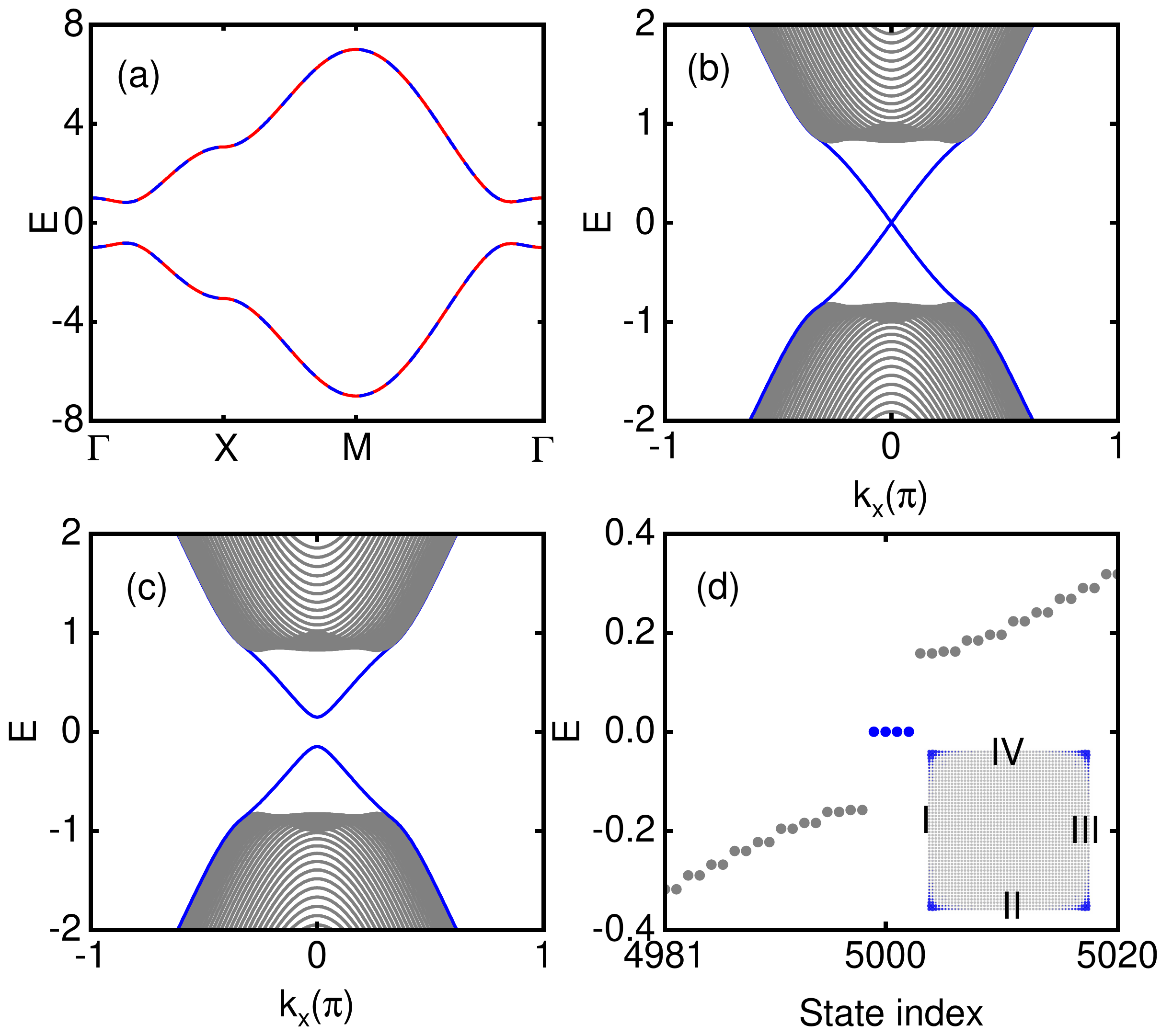}
\caption{
(a) Bulk band structure along the high-symmetry path for $J=0.3$. 
(b) and (c) Band structures of nanoribbons of the topological-insulator film for $J=0$ (b) and $J=0.3$ (c), respectively. Blue lines denote the edge states. 
(d) Energy levels of a rectangular nanoflake with the same parameters as those in panel (c). 
Blue dots indicate the in-gap corner states. 
Inset: Probability distribution of the four corner states.
The other parameters are chosen as $\upsilon_F=1$, $m_0=-1$, and $m_1=1$. The ribbon width is $N_y=50a$, and the nanoflake size is $50a\times 50a$.}
  \label{fig2}
\end{figure}

We first examine the boundary spectra of the Hamiltonian in Eq.~(\ref{eq:Hk_total}), as shown in Fig.~\ref{fig2}. 
We consider a nanoribbon geometry with periodic boundary conditions along the $x$ direction and open boundaries along the $y$ direction. 
The ribbon width is set to $N_y = 50a$, where $a$ denotes the lattice constant. 
Unless otherwise specified, the parameters are chosen as $\upsilon_F = 1$, $m_0 = -1$, and $m_1 = 1$.

Figure~\ref{fig2}(b) displays the ribbon spectrum of the parent topological-insulator film at $J = 0$. 
The bulk spectrum is fully gapped, while a pair of gapless helical edge states traverses the bulk gap and connects the valence and conduction bands, consistent with the quantum spin Hall phase. After introducing the compensated out-of-plane $d$-wave altermagnetic term, the helical edge states acquire a finite energy gap, as shown in Fig.~\ref{fig2}(c) for $J = 0.3$. This gap opening indicates that the altermagnetic perturbation acts as an effective Dirac mass for the boundary modes while preserving bulk $\mathcal{PT}$ symmetry. To further identify the higher-order boundary states, we next consider a finite square nanoflake with dimensions $50a \times 50a$. 
The corresponding energy spectrum is presented in Fig.~\ref{fig2}(d), where four in-gap states emerge precisely at zero energy. 
The inset of Fig.~\ref{fig2}(d) shows that these states are strongly localized at the four corners of the sample, providing clear evidence for the formation of zero-energy corner modes in the finite system.
As a result, the compensated altermagnetic term gaps the helical edge states and drives the topological-insulator film into a two-dimensional second-order topological insulating phase with corner states, while preserving bulk band degeneracy without spin splitting.

\subsection{Bulk band topology and edge theory}

We now turn to the bulk topological characterization of the compensated altermagnetic phase. Although the compensated altermagnetic term gaps the helical edge states of the parent quantum spin Hall film, it does not trivialize the system. The relevant topological information can be extracted from the two diagonal mirror-invariant lines, $k_x=k_y$ and $k_x=-k_y$, on which the Bloch Hamiltonian remains symmetric under the corresponding mirror operations. As a result, the Hamiltonian restricted to each of these lines can be decomposed into two mirror subspaces, in which mirror-resolved winding numbers can be defined.

\begin{figure} 
\centering 
\includegraphics[width=8.5cm,angle=0]{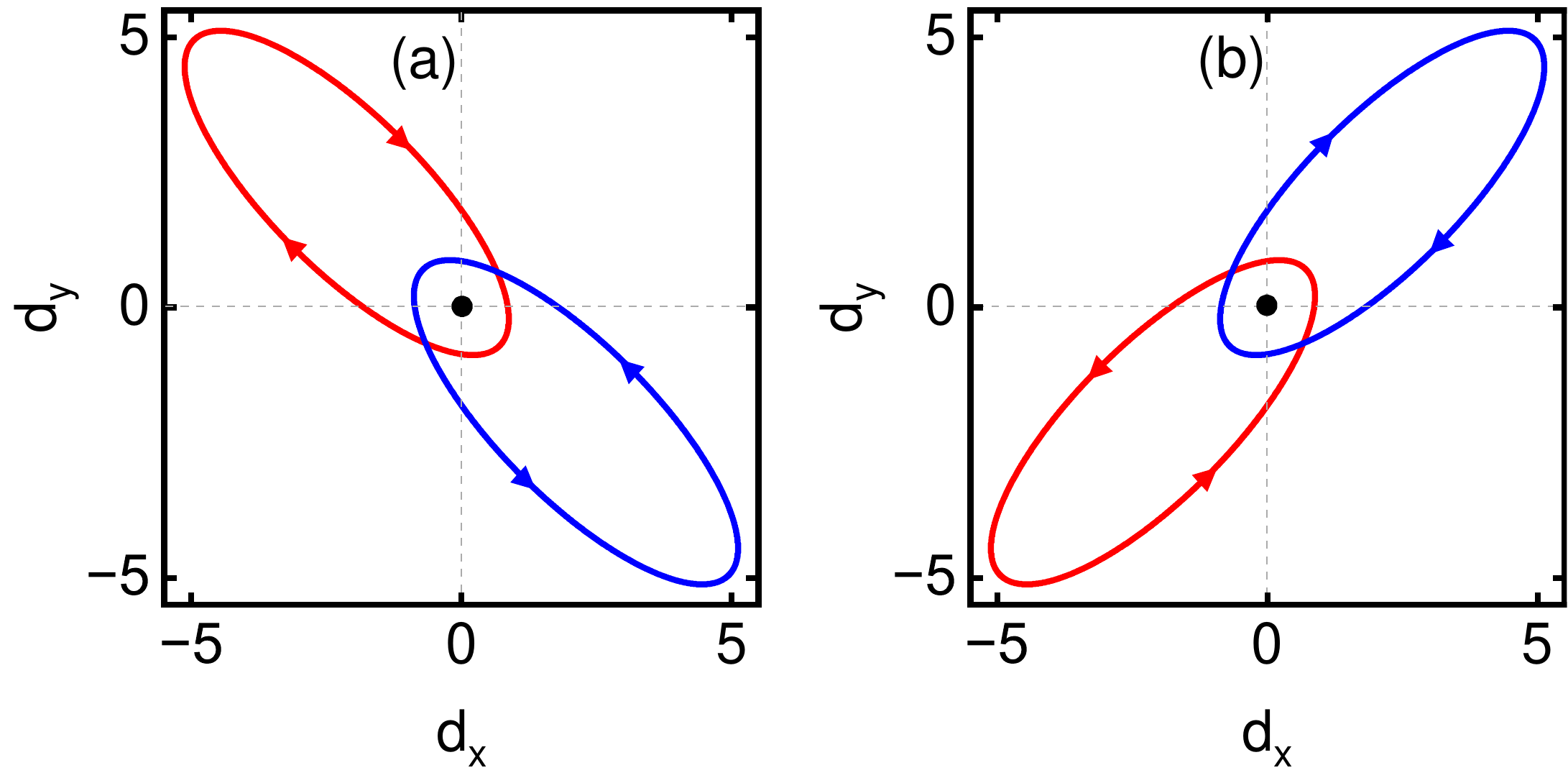} 
\caption{Winding trajectories of the effective Hamiltonians in the two mirror subspaces. (a) Along $k_x=k_y$. (b) Along $k_x=-k_y$. The red and blue curves denote the two mirror sectors with eigenvalues $\pm i$, respectively. }
 \label{fig3}
  \end{figure}

Along the diagonal line $k_x=k_y$, the Bloch Hamiltonian is invariant under the mirror operation
\begin{align}
\mathcal{M}_{x\bar{y}}= i\frac{\sqrt{2}}{2}\tau_x(s_x-s_y),
\label{eq:Mxmy}
\end{align}
with $\mathcal{M}_{x\bar{y}}^2=-1$. The Hamiltonian restricted to this line can therefore be block diagonalized into two mirror subspaces with eigenvalues $\pm i$. In these two sectors, the effective one-dimensional Hamiltonians take the form
\begin{align}
H_{\pm i}^{(1)}(k_x)=d_x^{\pm}(k_x)\sigma_x+d_y^{\pm}(k_x)\sigma_y,
\label{eq:Hpm1}
\end{align}
where
\begin{align}
d_x^{\pm}(k_x)&=\upsilon_F\sin k_x \pm \frac{\sqrt{2}}{2}\left[m_0+4m_1(1-\cos k_x)\right],
\nonumber\\
d_y^{\pm}(k_x)&=\upsilon_F\sin k_x \mp \frac{\sqrt{2}}{2}\left[m_0+4m_1(1-\cos k_x)\right].
\label{eq:dvec1}
\end{align}
Since each mirror block contains only $\sigma_x$ and $\sigma_y$, it possesses an effective chiral symmetry and is therefore characterized by an integer winding number. Equivalently, one may examine the trajectory of the vector $\mathbf{d}_{\pm}(k_x)=[d_x^{\pm}(k_x),d_y^{\pm}(k_x)]$ in the $(d_x,d_y)$ plane. As shown in Fig.~\ref{fig3}(a), the red and blue curves, corresponding to the $+i$ and $-i$ mirror sectors, respectively, wind around the origin with opposite orientations, yielding $\nu_{+i}^{x\bar y}=1$ and $\nu_{-i}^{x\bar y}=-1$.

Similarly, along the diagonal line $k_x=-k_y$, the Bloch Hamiltonian is invariant under
\begin{align}
\mathcal{M}_{xy}= i\frac{\sqrt{2}}{2}\tau_x(s_x+s_y),
\label{eq:Mxy}
\end{align}
and can be decomposed into two mirror subspaces. The corresponding effective Hamiltonians are
\begin{align}
H_{\pm i}^{(2)}(k_x)=\tilde d_x^{\pm}(k_x)\sigma_x+\tilde d_y^{\pm}(k_x)\sigma_y,
\label{eq:Hpm2}
\end{align}
with
\begin{align}
\tilde d_x^{\pm}(k_x)&=\upsilon_F\sin k_x \pm \frac{\sqrt{2}}{2}\left[m_0+4m_1(1-\cos k_x)\right],
\nonumber\\
\tilde d_y^{\pm}(k_x)&=-(\upsilon_F\sin k_x \mp \frac{\sqrt{2}}{2}\left[m_0+4m_1(1-\cos k_x)\right]).
\label{eq:dvec2}
\end{align}
The corresponding winding trajectories are plotted in Fig.~\ref{fig3}(b), giving $\nu_{+i}^{xy}=-1$ and $\nu_{-i}^{xy}=1$.

Therefore, both diagonal mirror-invariant lines carry nontrivial mirror-resolved winding structures. More explicitly, the corresponding mirror-graded winding numbers are
\begin{align}
\nu_M^{x\bar y}=\frac{\nu_{+i}^{x\bar y}-\nu_{-i}^{x\bar y}}{2}=1,
\qquad
\nu_M^{xy}=\frac{\nu_{+i}^{xy}-\nu_{-i}^{xy}}{2}=-1.
\label{eq:nuM}
\end{align}
These nonzero mirror-graded winding numbers, defined on the two diagonal mirror-invariant lines, reveal the nontrivial mirror-resolved topology inherited from the parent quantum spin Hall phase. Although the compensated altermagnetic term gaps the first-order helical edge states, it does not trivialize the system. 
Instead, it drives the system into a second-order topological insulating phase with corner states.

To further clarify how the compensated altermagnetic term converts the parent quantum spin Hall phase into a second-order one and to reveal the physical origin of the corner states, we next derive the effective edge theory. 
Expanding Eq.~(\ref{eq:Hk_total}) around the $\Gamma$ point, we obtain the continuum Hamiltonian
\begin{align}
H(\mathbf{k})
=& \upsilon_F(k_y s_x\tau_z-k_x s_y\tau_z)
+\left[m_0+m_1(k_x^2+k_y^2)\right]\tau_x
\nonumber\\
&-\frac{J}{2}(k_x^2-k_y^2)s_z\tau_z .
\label{eq:Hcont}
\end{align}
Although the compensated altermagnetic term preserves $\mathcal{PT}$ symmetry and therefore keeps the bulk bands doubly degenerate, it can still generate an effective Dirac mass for the helical edge states. We first consider edge $I$ in the inset of Fig.~\ref{fig2}(c), corresponding to the semi-infinite geometry $x\ge 0$. Replacing $k_x\rightarrow -i\partial_x$, and keeping only the leading-order terms in the edge momentum $k_y$, the Hamiltonian can be separated into
\begin{align}
H_0(-i\partial_x)
=&\, i\upsilon_F\partial_x s_y\tau_z+(m_0-m_1\partial_x^2)\tau_x,
\nonumber\\
H_p(-i\partial_x,k_y)
=&\, \upsilon_F k_y s_x\tau_z+\frac{J}{2}\partial_x^2 s_z\tau_z ,
\label{eq:H0Hp}
\end{align}
where the quadratic terms proportional to $k_y^2$, have been neglected since they only give higher-order corrections to the low-energy edge theory near $k_y=0$.
For edge states localized near $x=0$, the boundary conditions are $\psi(0)=\psi(\infty)=0$. Solving $H_0\psi=0$, one obtains two zero-mode wave functions
\begin{align}
\psi_1(x)&=C\sin(\gamma_1 x)e^{\gamma_2 x}e^{ik_y y}\chi_1,
\nonumber\\
\psi_2(x)&=C\sin(\gamma_1 x)e^{\gamma_2 x}e^{ik_y y}\chi_2,
\label{eq:edgewf}
\end{align}
where
\begin{align}
\chi_1=\frac{\sqrt{2}}{2}[1,0,0,-1]^T,\qquad
\chi_2=\frac{\sqrt{2}}{2}[0,1,1,0]^T,
\label{eq:chi12}
\end{align}
and
\begin{align}
\gamma_1=\sqrt{\left|m_0/m_1\right|-\upsilon_F^2/(2m_1)^2},
\qquad
\gamma_2=-\frac{\upsilon_F}{2m_1}.
\label{eq:gamma12}
\end{align}
Projecting $H_p$ onto the subspace spanned by the two edge zero modes, we obtain the effective Hamiltonian on edge $I$,
\begin{align}
H_I=\upsilon_F k_\parallel \eta_x+\frac{Jm_0}{2m_1}\eta_z,
\label{eq:HI}
\end{align}
where $k_\parallel$ denotes the momentum along the corresponding edge and $\eta_i$ are Pauli matrices in the basis of the two  subspaces.

Applying the same procedure to the other three edges yields
\begin{align}
H_{II}&=-\upsilon_F k_\parallel\eta_y-\frac{Jm_0}{2m_1}\eta_z,
\nonumber\\
H_{III}&=-\upsilon_F k_\parallel\eta_x+\frac{Jm_0}{2m_1}\eta_z,
\nonumber\\
H_{IV}&=\upsilon_F k_\parallel\eta_y-\frac{Jm_0}{2m_1}\eta_z .
\label{eq:Hedges}
\end{align}
Therefore, the Dirac masses on the four edges are
\begin{align}
m_I=\frac{Jm_0}{2m_1},\qquad
m_{II}=-\frac{Jm_0}{2m_1},
\nonumber\\
m_{III}=\frac{Jm_0}{2m_1},\qquad
m_{IV}=-\frac{Jm_0}{2m_1}.
\label{eq:masses}
\end{align}
One thus finds that adjacent edges carry Dirac masses with opposite signs. As a result, Dirac mass domain walls are formed at all four corners, and according to the Jackiw-Rebbi mechanism~\cite{Jackiw1976}, these domain walls bind localized zero-dimensional states. This edge-theory analysis provides a direct boundary interpretation of the higher-order topology and is fully consistent with the numerical corner states shown in Fig.~\ref{fig2}.


\subsection{Topological phase diagram}

\begin{figure}[t]
  \centering
  \includegraphics[width=8.5cm,angle=0]{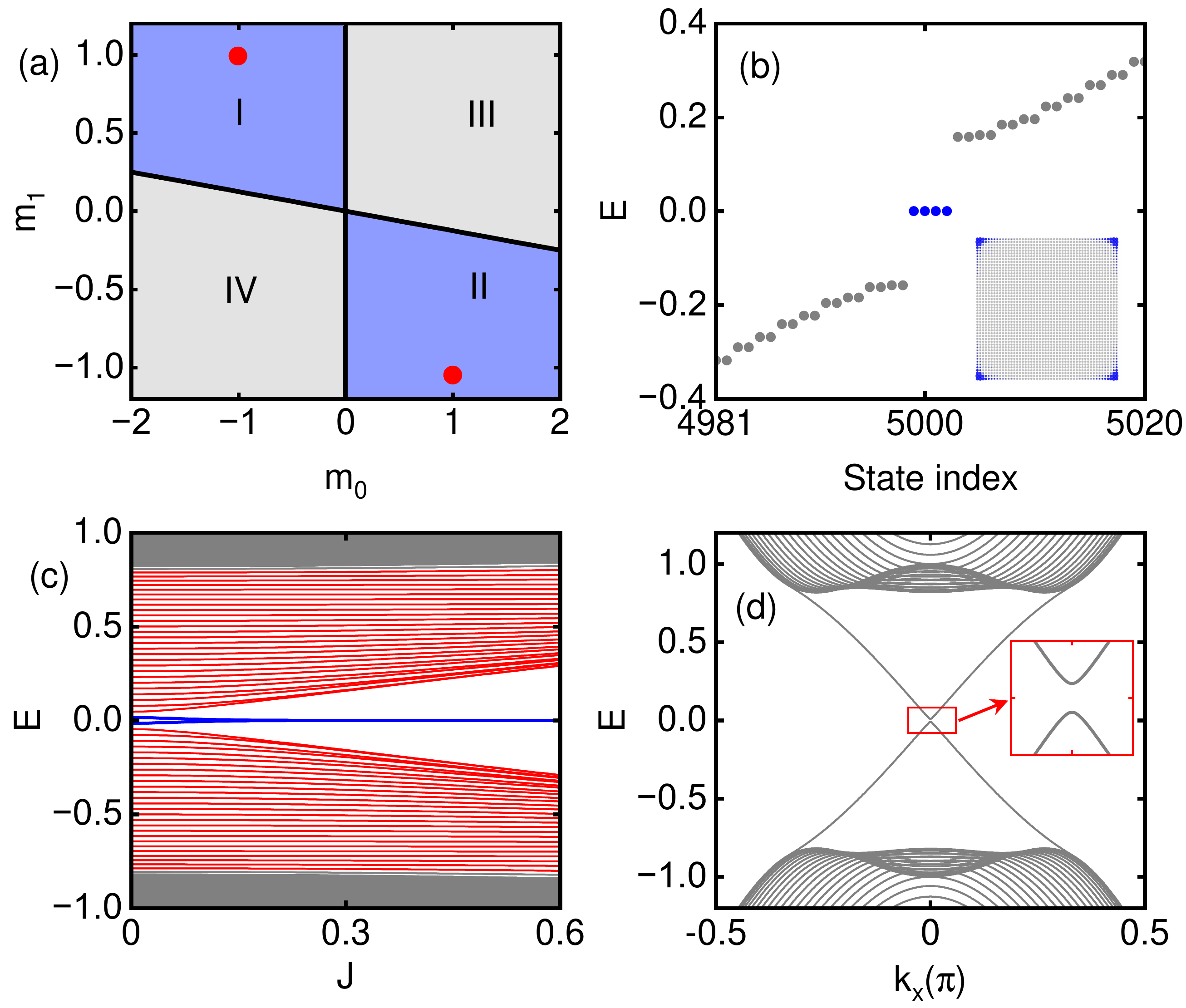}
\caption{
(a) Bulk phase diagram in the $(m_0,m_1)$ plane. 
The blue regions denote the second-order topological insulator, while the gray regions denote the trivial insulator. 
The red dots mark representative parameter points for regions I and II, given by $(m_0,m_1)=(-1,1)$ and $(1,-1)$, respectively.
(b) Energy spectrum of a square nanoflake for the representative parameter point chosen from region II. The inset shows the real-space probability distribution of the zero-energy states.
(c) Energy spectrum of a nanoflake as a function of $J$. Gray lines denote bulk levels, red lines highlight boundary levels, and blue lines denote the four zero-energy states.
(d) Nanoribbon spectrum for $J=0.02$. The inset gives an enlarged view near zero energy.
The nanoflake size in panels (b) and (c) is $N_x=N_y=50$. The other parameters are $\upsilon_F=1$ and $J=0.3$ in panels (a) and (b), while $\upsilon_F=1$, $m_0=-1$, and $m_1=1$ in panels (c) and (d).}
  \label{fig4}
\end{figure}

To determine the parameter regime of the compensated-altermagnetism-induced second-order topological phase, we first analyze the bulk gap-closing conditions of Eq.~(\ref{eq:Hk_total}), which determine the bulk phase boundaries in the $(m_0,m_1)$ plane. Defining
\begin{align}
M(\mathbf{k})&=m_0+2m_1(2-\cos k_x-\cos k_y),
\nonumber\\
\Delta(\mathbf{k})&=\cos k_x-\cos k_y,
\label{eq:MD_def}
\end{align}
the bulk spectrum can be written as
\begin{align}
E(\mathbf{k})=\pm\sqrt{\upsilon_F^2(\sin^2 k_x+\sin^2 k_y)+M^2(\mathbf{k})+J^2\Delta^2(\mathbf{k}) }.
\label{eq:bulk_spectrum_phase}
\end{align}
A bulk gap closing requires $\sin k_x=\sin k_y=0$, $M(\mathbf{k})=0$, and $\Delta(\mathbf{k})=0$ to be satisfied simultaneously. 
This condition can only be satisfied at $\Gamma=(0,0)$ and $M=(\pi,\pi)$, where the compensated $d$-wave altermagnetic term vanishes. 
This again indicates that the magnitude of the compensated altermagnetic term does not affect the bulk band gap.
The corresponding bulk phase boundaries are therefore
\begin{align}
m_0=0,
\qquad
m_0+8m_1=0.
\label{eq:phase_boundary_fig4}
\end{align}
These bulk phase boundaries separate the parent quantum spin Hall regime from the trivial insulating regime. Within the former, turning on a finite compensated altermagnetic term does not close the bulk gap, but gaps the helical edge states and converts the parent first-order topological phase into a second-order one.
We therefore obtain the bulk phase diagram in the $(m_0,m_1)$ plane at fixed $J=0.3$, as shown in Fig.~\ref{fig4}(a). The red dots mark the representative parameter points used in the calculations.

Region I, corresponding to the parameter set $(m_0,m_1)=(-1,1)$ used in Fig.~\ref{fig2}, has already been shown to host a second-order topological insulating phase with four corner states. We further calculate the open-boundary energy spectrum for a representative parameter point chosen from region II, namely $(m_0,m_1)=(1,-1)$, as shown in Fig.~\ref{fig4}(b).

For region II, Fig.~\ref{fig4}(b) shows four zero-energy in-gap states. The inset further confirms that these states are localized at the four corners of the finite sample, demonstrating that region II also belongs to the second-order topological insulating phase. 
Indeed, regions I and II are exactly equivalent in the present model, since the Hamiltonian satisfies $\tau_z H(\mathbf{k};m_0,m_1,J)\tau_z = H(\mathbf{k};-m_0,-m_1,J)$. They therefore represent the same second-order topological insulating phase rather than two distinct ones. By contrast, regions III and IV, which lie in the bulk-trivial regime, are topologically trivial.

We emphasize that the compensated $d$-wave altermagnetic term does not shift the bulk phase boundaries in the $(m_0,m_1)$ plane, since it vanishes at the bulk gap-closing momenta $\Gamma$ and $M$. Its role is instead to generate a boundary Dirac mass. From the effective edge theory in Eq.~(\ref{eq:masses}), the edge masses are proportional to \(J\), and therefore the induced boundary gap is proportional to \(|J|\). Thus, within the inverted bulk regime, there is no finite critical value of $J$ for opening the edge gap. The only boundary gap-closing point is $J=0$, and any finite $J$ generates a nonzero boundary mass in the thermodynamic edge theory. This behavior is confirmed by the finite-nanoflake spectrum in Fig.~\ref{fig4}(c). As $J$ increases from zero, the bulk levels remain gapped, while the boundary levels are continuously gapped, and four zero-energy corner states remain inside the induced boundary gap. The nanoribbon spectrum at a small finite value $J=0.02$, shown in Fig.~\ref{fig4}(d), further confirms that the helical edge states are already gapped once $J\neq 0$. Thus, the compensated altermagnetic strength $J$ drives the higher-order phase by opening the boundary gap, while leaving the bulk phase-transition points unchanged.

\subsection{Robustness against symmetry-breaking perturbations}

We next examine the stability of the corner states against representative symmetry-breaking perturbations. 
Since the ideal phase relies on \(\mathcal{PT}\) symmetry and mirror-resolved topology, the key issue is whether weak symmetry breaking destroys the corner states before the bulk or boundary gap closes.

\begin{figure}[t]
  \centering
  \includegraphics[width=8.5cm,angle=0]{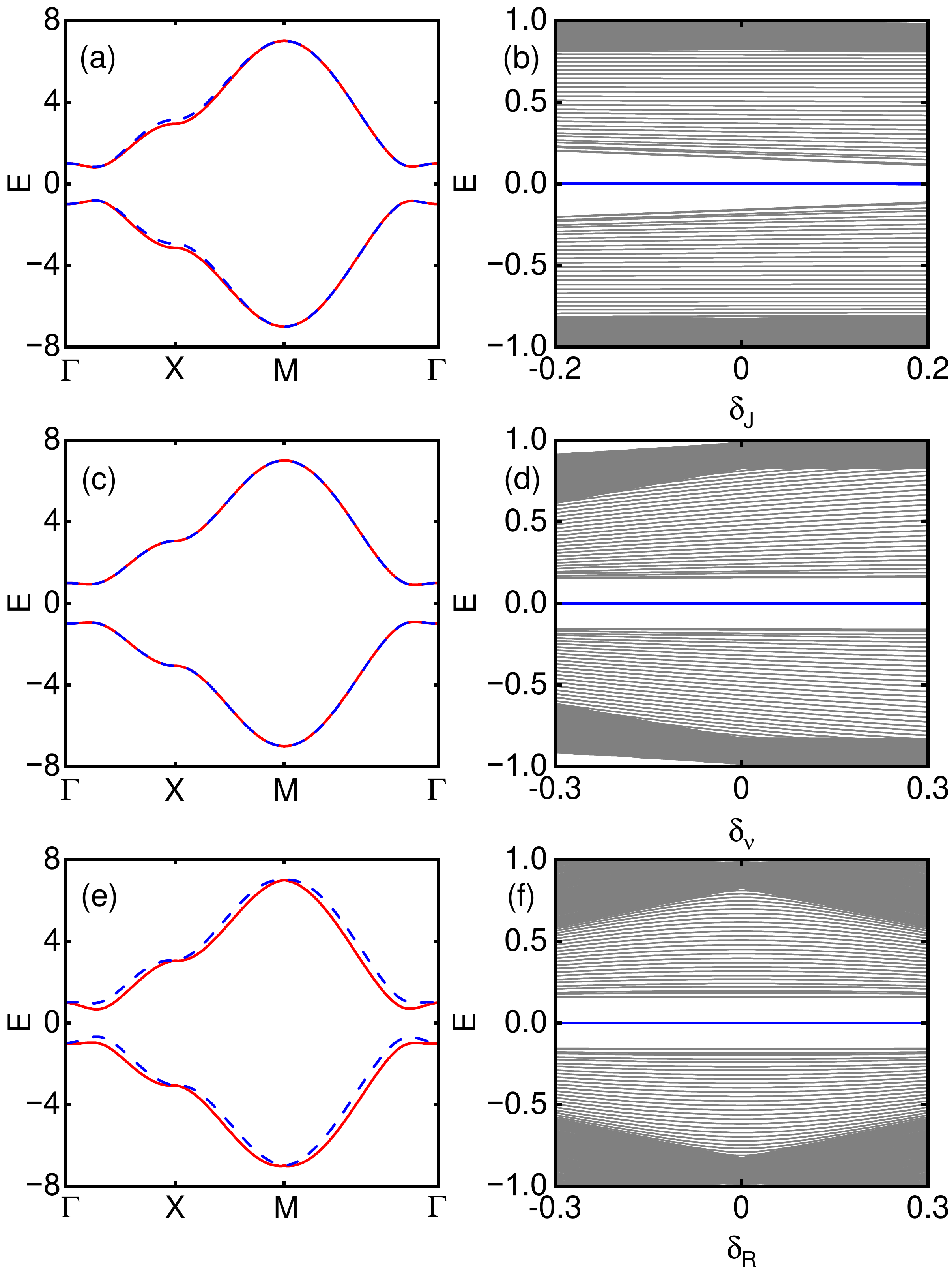}
\caption{
(a), (c), and (e) Bulk band structures for \(\delta_J=0.1\), \(\delta_v=0.2\), and \(\delta_R=0.2\), respectively. Red and blue curves distinguish the two branches of each bulk band.
(b), (d), and (f) Energy spectra of a nanoflake as functions of \(\delta_J\), \(\delta_v\), and \(\delta_R\), respectively. The blue lines denote the four zero-energy states.
The unperturbed parameters are \(J=0.3\), \(N_x=N_y=50\), \(v_F=1\), \(m_0=-1\), and \(m_1=1\).
}  \label{fig5}
\end{figure}

We first consider an imbalance between the altermagnetic exchange strengths on the two layers. Keeping the altermagnetic strength on the top layer fixed as \(J\), while taking that on the bottom layer to be \(J-\delta_J\) with the opposite sign, gives
\begin{align}
H_{\rm AM}^{\rm im}(k)
=
\Delta (\mathbf{k})s_z
\left[
\frac{\delta_J}{2}\tau_0+
\left(J-\frac{\delta_J}{2}\right)\tau_z
\right],
\label{eq:imbalance_AM}
\end{align}
where \(\Delta(\mathbf{k})=\cos k_x-\cos k_y\). A finite \(\delta_J\) breaks the exact layer compensation and lifts the \(\mathcal{PT}\)-protected bulk double degeneracy. As shown in Fig.~\ref{fig5}(a), the bulk bands are split for \(\delta_J=0.1\), while no bulk gap closing occurs. Correspondingly, Fig.~\ref{fig5}(b) shows that the four zero-energy states remain inside the boundary gap for weak \(\delta_J\).

Next, we consider mirror-symmetry breaking by introducing an in-plane velocity anisotropy. Specifically, we change the Dirac velocity along the \(x\) direction, which corresponds to adding the perturbation
\begin{align}
\delta H_v(\mathbf{k})
=
-\delta_v\sin k_x s_y\tau_z .
\label{eq:velocity_anisotropy_delta}
\end{align}
This perturbation makes the two in-plane Dirac velocities unequal and therefore breaks the diagonal mirror symmetries. Consequently, the Hamiltonian can no longer be block diagonalized into mirror subspaces, so the mirror-graded winding numbers are not well defined for finite \(\delta_v\).
However, this perturbation does not break the layer-opposite \(\mathcal{PT}\)-symmetric structure. As shown in Fig.~\ref{fig5}(c), the bulk gap remains open, and the bulk bands remain doubly degenerate for \(\delta_v=0.2\), without observable band splitting. The corresponding open-boundary spectrum in Fig.~\ref{fig5}(d) shows that the four zero-energy states remain separated from the other states.

Finally, we examine Rashba spin-orbit coupling,
\begin{align}
H_R(\mathbf{k})
=
\delta_R
\left(
\sin k_y s_x
-
\sin k_x s_y
\right)\tau_0 ,
\label{eq:rashba_perturbation}
\end{align}
where \(\delta_R\) is the Rashba spin-orbit coupling strength. This term mixes the spin and mirror sectors and breaks the ideal \(\mathcal{PT}\)-symmetric structure of the compensated model. Accordingly, the bulk double degeneracy is lifted, as shown in Fig.~\ref{fig5}(e) for \(\delta_R=0.2\). Nevertheless, the bulk gap remains open, and the four zero-energy states persist in the finite open-boundary spectrum shown in Fig.~\ref{fig5}(f).

We have also checked the real-space probability distributions of the four zero-energy states under these perturbations and confirmed that they remain localized near the sample corners, similar to the corner-state distributions shown in the inset of Fig.~\ref{fig2}(d). 
Therefore, although exact symmetry protection is guaranteed only in the ideal \(\mathcal{PT}\)- and mirror-symmetric limit, the corner-state signatures remain stable against weak layer imbalance, weak mirror-symmetry breaking, and weak Rashba spin-orbit coupling, provided that the bulk and boundary gaps remain open.

\section{Conclusions}

In summary, we have shown that compensated altermagnetism can drive a topological-insulator film into a two-dimensional second-order topological insulating phase without affecting the bulk band gap. By introducing a layer-resolved out-of-plane $d$-wave altermagnetic term with opposite signs in the top and bottom layers, we obtain a compensated altermagnetic background that preserves $\mathcal{PT}$ symmetry and therefore exhibits no bulk spin splitting. Nevertheless, this term gaps the helical edge states of the parent topological-insulator film and gives rise to zero-energy corner states in finite samples.

We have further shown that the resulting higher-order phase is characterized by nonzero mirror-graded winding numbers along the two diagonal high-symmetry lines. The effective edge theory demonstrates that the compensated altermagnetic term generates Dirac masses with alternating signs on adjacent edges, so that the corner states originate from the corresponding mass domain walls. In addition, the phase diagram confirms that this second-order topological phase occupies a finite parameter region.

Our results identify compensated altermagnetism as an effective route to engineering higher-order topology in a $\mathcal{PT}$-symmetric setting without bulk spin splitting. From an experimental viewpoint, a possible route is to use Bi$_2$Se$_3$ topological-insulator films as the platform, since high-quality thin-film growth, magnetic proximity coupling, and magnetic heterostructure engineering have already been extensively developed in Bi$_2$Se$_3$-based systems~\cite{Zhang2009, Zhu2018,Fanchiang2018,Liu2020,Wang2023}. The required $d$-wave momentum dependence may be supplied by altermagnetic layers whose exchange field has crystalline anisotropy, as indicated by recent studies of metallic and layered insulating $d$-wave altermagnets~\cite{Jiang2025, Wei2025}. More specifically, the layer-opposite $d$-wave altermagnetic term considered here may be realized when the top and bottom surfaces of the topological-insulator film are proximity-coupled to $d$-wave altermagnetic layers with opposite altermagnetic orientations or opposite interface terminations, thereby inducing opposite altermagnetic exchange fields on the two surfaces. This physical picture is further supported by theoretical work showing that momentum-dependent altermagnetic spin splitting can be transferred to adjacent nonmagnetic layers through interfacial coupling~\cite{Zhu2026}. In addition, recent experimental progress on altermagnetic thin films and layered altermagnets suggests that incorporating altermagnetic order into heterostructures is becoming increasingly feasible~\cite{Krempasky2024, Reimers2024, Regmi2025}.
Therefore, the required ingredients are compatible with currently developing topological-insulator/altermagnet heterostructure platforms.

\section*{Acknowledgements}

This work was financially supported by the National Natural Science Foundation of China (Grants No. 12074097, No. 12374034, and No. 12547169),
Natural Science Foundation of Hebei Province (Grant No. A2024205025),
the National Key R and D Program of China (Grant No. 2024YFA1409002),
and the Quantum Science and Technology-National Science and Technology Major Project (Grant No. 2021ZD0302403).


\begin{thebibliography}{99}
\bibitem{Haldane1988} F. D. M. Haldane, Model for a Quantum Hall Effect without Landau Levels: Condensed-Matter Realization of the "Parity Anomaly", Phys. Rev. Lett. \textbf{61}, 2015 (1988).

\bibitem{Kane2005} C. L. Kane and E. J. Mele, ${Z}_{2}$ Topological Order and the Quantum Spin Hall Effect, Phys. Rev. Lett. \textbf{95}, 146802 (2005).
\bibitem{Kane2005a} C. L. Kane and E. J. Mele, Quantum Spin Hall Effect in Graphene, Phys. Rev. Lett. \textbf{95}, 226801 (2005).
\bibitem{Bernevig2006} B. A. Bernevig, T. L. Hughes, and S.-C. Zhang, Quantum Spin Hall Effect and Topological Phase Transition in HgTe Quantum Wells, Science \textbf{314}, 1757 (2006).


\bibitem{Qi2011} X.-L. Qi and S.-C. Zhang, Topological insulators and superconductors, Rev. Mod. Phys. \textbf{83}, 1057 (2011).

\bibitem{Hasan2010} M. Z. Hasan and C. L. Kane, Colloquium: Topological insulators, Rev. Mod. Phys. \textbf{82}, 3045 (2010).
\bibitem{Ren2016} Y. Ren, Z. Qiao, and Q. Niu, Topological phases in two-dimensional materials: a review, Rep. Prog. Phys. \textbf{79}, 066501 (2016).



\bibitem{Benalcazar2017} W. A. Benalcazar, B. A. Bernevig, and T. L. Hughes, Quantized electric multipole insulators, Science \textbf{357}, 61 (2017).
\bibitem{Benalcazar2017a} W. A. Benalcazar, B. A. Bernevig, and T. L. Hughes, Electric multipole moments, topological multipole moment pumping, and chiral hinge states in crystalline insulators, Phys. Rev. B \textbf{96}, 245115 (2017).

    \bibitem{Schindler2018} F. Schindler, A. M. Cook, M. G. Vergniory, Z. Wang, S. S. P. Parkin, B. A. Bernevig, and T. Neupert, Higher-order topological insulators, Sci. Adv. \textbf{4}, eaat0346 (2018).

\bibitem{Li2020} T. Li, P. Zhu, W. A. Benalcazar, and T. L. Hughes, Fractional disclination charge in two-dimensional ${C}_{n}$-symmetric topological crystalline insulators, Phys. Rev. B \textbf{101}, 115115 (2020).



\bibitem{Benalcazar2019} W. A. Benalcazar, T. Li, and T. L. Hughes, Quantization of fractional corner charge in ${C}_{n}$-symmetric higher-order topological crystalline insulators, Phys. Rev. B \textbf{99}, 245151 (2019).

\bibitem{Langbehn2017} J. Langbehn, Y. Peng, L. Trifunovic, F. von Oppen, and P. W. Brouwer, Reflection-Symmetric Second-Order Topological Insulators and Superconductors, Phys. Rev. Lett. \textbf{119}, 246401 (2017).

\bibitem{Schindler2018a} F. Schindler, Z. Wang, M. G. Vergniory, A. M. Cook, A. Murani, S. Sengupta, A. Y. Kasumov, R. Deblock, S. Jeon, I. Drozdov, H. Bouchiat, S. Guron, A. Yazdani, B. A. Bernevig, and T. Neupert, Higher-order topology in bismuth, Nat. Phys. \textbf{14}, 918 (2018).

\bibitem{Peterson2018} C. W. Peterson, W. A. Benalcazar, T. L. Hughes, and G. Bahl, A quantized microwave quadrupole insulator with topologically protected corner states, Nature \textbf{555}, 346 (2018).

\bibitem{Xu2019} Y. Xu, Z. Song, Z. Wang, H. Weng, and X. Dai, Higher-Order Topology of the Axion Insulator ${\mathrm{EuIn}}_{2}{\mathrm{As}}_{2}$, Phys. Rev. Lett. \textbf{122}, 256402 (2019).


\bibitem{Chen2020a} R. Chen, C.-Z. Chen, J.-H. Gao, B. Zhou, and D.-H. Xu, Higher-Order Topological Insulators in Quasicrystals, Phys. Rev. Lett. \textbf{124}, 036803 (2020).


\bibitem{Liu2024} L. Liu, C.-M. Miao, Q.-F. Sun, and Y.-T. Zhang, Two-dimensional higher-order topological metals, Phys. Rev. B \textbf{110}, 205415 (2024).

\bibitem{Liu2024b} L. Liu, C. Miao, H. Tang, Y.-T. Zhang, and Z. Qiao, Magnetically controlled topological braiding with Majorana corner states in second-order topological superconductors, Phys. Rev. B \textbf{109}, 115413 (2024).

    \bibitem{Liu2019} B. Liu, G. Zhao, Z. Liu, and Z. F. Wang, Two-Dimensional Quadrupole Topological Insulator in \textbf{$\gamma$}-Graphyne, Nano Lett. \textbf{19}, 6492 (2019).




\bibitem{Park2019} M. J. Park, Y. Kim, G. Y. Cho, and S. Lee, Higher-Order Topological Insulator in Twisted Bilayer Graphene, Phys. Rev. Lett. \textbf{123}, 216803 (2019).



\bibitem{Lee2020} E. Lee, R. Kim, J. Ahn, and B.-J. Yang, Two-dimensional higher-order topology in monolayer graphdiyne, npj Quantum Mater. \textbf{5}, 1 (2020).

\bibitem{Sheng2019} X.-L. Sheng, C. Chen, H. Liu, Z. Chen, Z.-M. Yu, Y. Zhao, and S. A. Yang, Two-Dimensional Second-Order Topological Insulator in Graphdiyne, Phys. Rev. Lett. \textbf{123}, 256402 (2019).

 \bibitem{Liu2024a} L. Liu, J. An, Y. Ren, Y.-T. Zhang, Z. Qiao, and Q. Niu, Engineering second-order topological insulators via coupling two first-order topological insulators, Phys. Rev. B \textbf{110}, 115427 (2024).


\bibitem{Miao2023} C.-M. Miao, Y.-H. Wan, Q.-F. Sun, and Y.-T. Zhang, Engineering topologically protected zero-dimensional interface end states in antiferromagnetic heterojunction graphene nanoflakes, Phys. Rev. B \textbf{108}, 075401 (2023).

 
 
 

\bibitem{Xie2019} B. Xie, G. Su, H. Wang, H. Su, X. Shen, P. Zhan, M. Lu, Z. Wang, and Y. Chen, Visualization of Higher-Order Topological Insulating Phases in Two-Dimensional Dielectric Photonic Crystals, Phys. Rev. Lett. \textbf{122}, 233903 (2019).
\bibitem{Zhu2021} B. Zhu, Q. Wang, Y. Zeng, Q. J. Wang, and Y. D. Chong, Single-mode lasing based on $\mathcal{PT}$-breaking of two-dimensional photonic higher-order topological insulator, Phys. Rev. B \textbf{104}, L140306 (2021).
\bibitem{Zhu2022} J. Zhu, W. Wu, J. Zhao, C. Chen, Q. Wang, X.-L. Sheng, L. Zhang, Y. X. Zhao, and S. A. Yang, Phononic real Chern insulator with protected corner modes in graphynes, Phys. Rev. B \textbf{105}, 085123 (2022).
\bibitem{Huang2023} F. Huang, P. Zhou, W. Li, S. He, R. Tan, Z. Ma, and L. Z. Sun, Phononic second-order topological phase in the ${\mathrm{C}}_{3}\mathrm{N}$ compound, Phys. Rev. B \textbf{107}, 134104 (2023).
\bibitem{Han2024} Y. Han, C. Cui, X.-P. Li, T.-T. Zhang, Z. Zhang, Z.-M. Yu, and Y. Yao, Cornertronics in Two-Dimensional Second-Order Topological Insulators, Phys. Rev. Lett. \textbf{133}, 176602 (2024).



\bibitem{Ren2020} Y. Ren, Z. Qiao, and Q. Niu, Engineering Corner States from Two-Dimensional Topological Insulators, Phys. Rev. Lett. \textbf{124}, 166804 (2020).

\bibitem{Chen2020} C. Chen, Z. Song, J.-Z. Zhao, Z. Chen, Z.-M. Yu, X.-L. Sheng, and S. A. Yang, Universal Approach to Magnetic Second-Order Topological Insulator, Phys. Rev. Lett. \textbf{125}, 056402 (2020).

\bibitem{Miao2024} C.-M. Miao, L. Liu, Y.-H. Wan, Q.-F. Sun, and Y.-T. Zhang, General principle behind magnetization-induced second-order topological corner states in the Kane-Mele model, Phys. Rev. B \textbf{109}, 205417 (2024).

\bibitem{Li2024} Y.-X. Li, Y. Liu and C.-C. Liu, Creation and manipulation of higher-order topological states by altermagnets, Phys. Rev. B \textbf{109}, L201109 (2024).

\bibitem{Miao2025} C.-M. Miao, Y.-H. Wan, Y.-T. Zhang, and Q.-F. Sun, Tunable second-order topological corner states induced by interlayer coupling in twisted bilayer Chern insulators, Phys. Rev. B \textbf{111}, 165418 (2025).

  \bibitem{Liu2025} L. Liu, J. An, Y. Ren, Y.-T. Zhang, Z. Qiao, and Q. Niu, Engineering corner states by coupling two-dimensional topological insulators, Phys. Rev. B \textbf{111}, 045403 (2025).




\bibitem{Smejkal2022} L. ${\rm \check{S}}$mejkal, J. Sinova, and T. Jungwirth, Emerging Research Landscape of Altermagnetism, Phys. Rev. X \textbf{12}, 040501 (2022).
\bibitem{Smejkal2022a} L. ${\rm \check{S}}$mejkal, A. H. MacDonald, J. Sinova, S. Nakatsuji, and T. Jungwirth, Anomalous Hall antiferromagnets, Nat. Rev. Mater. \textbf{7}, 482 (2022).

\bibitem{Ahn2019} K.-H. Ahn, A. Hariki, K.-W. Lee, and J. Kune${\rm \check{s}}$, Antiferromagnetism in ${\mathrm{RuO}}_{2}$ as $d$-wave Pomeranchuk instability, Phys. Rev. B \textbf{99}, 184432 (2019).

\bibitem{Smejkal2022b} L. ${\rm \check{S}}$mejkal, J. Sinova, and T. Jungwirth, Beyond Conventional Ferromagnetism and Antiferromagnetism: A Phase with Nonrelativistic Spin and Crystal Rotation Symmetry, Phys. Rev. X \textbf{12}, 031042 (2022).

\bibitem{Yuan2020} L.-D. Yuan, Z. Wang, J.-W. Luo, E. I. Rashba, and A. Zunger, Giant momentum-dependent spin splitting in centrosymmetric low-$Z$ antiferromagnets, Phys. Rev. B \textbf{102}, 014422 (2020).


\bibitem{Smejkal2020} L. ${\rm \check{S}}$mejkal, R. Gonz${\rm \acute{a}}$lez-Hern${\rm \acute{a}}$ndez, T. Jungwirth, and J. Sinova, Crystal time-reversal symmetry breaking and spontaneous Hall effect in collinear antiferromagnets, Sci. Adv. \textbf{6}, eaaz8809 (2020).


\bibitem{Betancourt2023} R. D. Gonzalez Betancourt, J. Zub${\rm \acute{a}}$${\rm \check{c}}$, R. Gonzalez-Hernandez, K. Geishendorf, Z. ${\rm \check{S}}$ob${\rm \acute{a}}$${\rm \check{n}}$, G. Springholz, K. Olejn${\rm \acute{l}}$k, L. ${\rm \check{S}}$mejkal, J. Sinova, T. Jungwirth, S. T. B. Goennenwein, A. Thomas, H. Reichlov${\rm \acute{a}}$, J. ${\rm \check{Z}}$elezn${\rm \acute{y}}$, and D. Kriegner, Spontaneous Anomalous Hall Effect Arising from an Unconventional Compensated Magnetic Phase in a Semiconductor, Phys. Rev. Lett. \textbf{130}, 036702 (2023).

\bibitem{Liu2025t} L. Liu, Q.-F. Sun, and Y.-T. Zhang, Tunable two-dimensional Dirac-Weyl semimetal phase induced by altermagnetism, Phys. Rev. B \textbf{112}, L161411 (2025).


\bibitem{Sun2023} C. Sun, A. Brataas, and J. Linder, Andreev reflection in altermagnets, Phys. Rev. B \textbf{108}, 054511 (2023).
\bibitem{Zhou2023} X. Zhou, W. Feng, R.-W. Zhang, L. Smejkal, J. Sinova, Y. Mokrousov, and Y. Yao, Crystal Thermal Transport in Altermagnetic ${\mathrm{RuO}}_{2}$, Phys. Rev. Lett. \textbf{132}, 056701 (2023).

\bibitem{Wan2025} Y.-H. Wan, C.-M. Miao, P.-Y. Liu, and Q.-F. Sun, Helical Fermi arc in altermagnetic Weyl semimetal, Phys. Rev. B \textbf{112}, 235411 (2025).
\bibitem{Zhang2023} S.-B. Zhang, L.-H. Hu, and T. Neupert, Finite-momentum Cooper pairing in proximitized altermagnets, Nat. Commun. \textbf{15}, 1801 (2023).




\bibitem{Yi2026} X.-J. Yi, Y. Mao, C.-M. Miao, and Q.-F. Sun, Majorana modes in a helical altermagnet without net magnetism and spin-orbit coupling, Phys. Rev. B \textbf{113}, L060408 (2026).

\bibitem{Sun2025a} H.-Y. Sun, L. Liu, Y.-T. Zhang, and Z. Qiao, Altermagnetism-induced iron-based third-order topological superconductivity, Phys. Rev. B \textbf{112}, 195402 (2025).



\bibitem{Cheng2024} Q. Cheng, Y. Mao, and Q.-F. Sun, Field-free Josephson diode effect in altermagnet/normal metal/altermagnet junctions, Phys. Rev. B \textbf{110}, 014518 (2024).


\bibitem{Liu2026q} L. Liu and Q.-F. Sun, Quantum anomalous Hall effect with tunable Chern numbers induced by d-wave sublattice-staggered altermagnetism, Chin. Phys. B \textbf{35}, 057301 (2026).

\bibitem{Sun2025t} Y.-F. Sun, Y. Mao, Y.-C. Zhuang, and Q.-F. Sun, Tunneling magnetoresistance effect in altermagnets, Phys. Rev. B \textbf{112}, 094411 (2025).


\bibitem{Zeng2024} S. Zeng and Y.-J. Zhao, Bilayer stacking $A$-type altermagnet: A general approach to generating two-dimensional altermagnetism, Phys. Rev. B \textbf{110}, 174410 (2024).
\bibitem{Pan2024} B. Pan, P. Zhou, P. Lyu, H. Xiao, X. Yang, and L. Sun, General Stacking Theory for Altermagnetism in Bilayer Systems, Phys. Rev. Lett. \textbf{133}, 166701 (2024).
\bibitem{Liu2024c} Y. Liu, J. Yu, and C.-C. Liu, Twisted Magnetic Van der Waals Bilayers: An Ideal Platform for Altermagnetism, Phys. Rev. Lett. \textbf{133}, 206702, (2024).

\bibitem{Meier2026} Q. N. Meier, A. Carta, C. Ederer, and A. Cano, Net and Compensated Altermagnetism from Staggered Orbital Order: Layer-Dependent Spin Splitting in Sr$_{n+1}$Cr$_n$O$_{3n+1}$, Phys. Rev. Lett. \textbf{136}, 116705 (2026).



\bibitem{Lu2010} H.-Z. Lu, W.-Y. Shan, W. Yao, Q. Niu, and S.-Q. Shen, Massive Dirac fermions and spin physics in an ultrathin film of topological insulator, Phys. Rev. B \textbf{81}, 115407 (2010).
\bibitem{Wang2014} J. Wang, B. Lian, and S.-C. Zhang, Universal scaling of the quantum anomalous Hall plateau transition, Phys. Rev. B \textbf{89}, 085106 (2014).

\bibitem{Shafiei2024} M. Shafiei, F. Fazileh, F. M. Peeters, and M. V. Milo\ifmmode \check{s}\else \v{s}\fi{}evi\ifmmode \acute{c}\else \'{c}\fi{}, Tailoring weak and metallic phases in a strong topological insulator by strain and disorder: Conductance fluctuations signatures, Phys. Rev. B \textbf{109}, 045129 (2024).

\bibitem{Li2020a} Q. Li, Y. Han, K. Zhang, Y.-T. Zhang, J.-J. Liu, and Z. Qiao, Multiple Majorana edge modes in magnetic topological insulator-superconductor heterostructures, Phys. Rev. B \textbf{102}, 205402 (2020).

\bibitem{Jiang2025} B. Jiang, M. Hu, J. Bai, Z. Song, C. Mu, G. Qu, W. Li, W. Zhu, H. Pi, Z. Wei, Y. Sun, Y. Huang, X. Zheng, Y. Peng, L. He, S. Li, J. Luo, Z. Li, G. Chen, H. Li, H. Weng, and T. Qian, A metallic room-temperature d-wave altermagnet, Nat. Phys. \textbf{21}, 754 (2025).


\bibitem{Jackiw1976} R. Jackiw and C. Rebbi, Solitons with fermion number 1/2, Phys. Rev. D \textbf{13}, 3398 (1976).

\bibitem{Zhang2009} H. Zhang, C.-X. Liu, X.-L. Qi, X. Dai, Z. Fang, and S.-C. Zhang, Topological insulators in Bi$_2$Se$_3$, Bi$_2$Te$_3$ and Sb$_2$Te$_3$ with a single Dirac cone on the surface, Nat. Phys. \textbf{5}, 438 (2009).

 \bibitem{Zhu2018} S. Zhu, D. Meng, G. Liang, G. Shi, P. Zhao, P. Cheng, Y. Li, X. Zhai, Y. Lu, L. Chen, and K. Wu, Proximity-induced magnetism and an anomalous Hall effect in Bi$_2$Se$_3$/LaCoO$_3$: a topological insulator/ferromagnetic insulator thin film heterostructure, Nanoscale \textbf{10}, 10041 (2018). 
 
 \bibitem{Fanchiang2018} Y. T. Fanchiang, K. H. M. Chen, C. C. Tseng, C. C. Chen, C. K. Cheng, S. R. Yang, C. N. Wu, S. F. Lee, M. Hong, and J. Kwo, Strongly exchange-coupled and surface-state-modulated magnetization dynamics in Bi$_2$Se$_3$/yttrium iron garnet heterostructures, Nat. Commun. \textbf{9}, 223 (2018).
 
 \bibitem{Liu2020} T. Liu, J. Kally, T. Pillsbury, C. Liu, H. Chang, J. Ding, Y. Cheng, M. Hilse, R. Engel-Herbert, A. Richardella, N. Samarth, and M. Wu, Changes of Magnetism in a Magnetic Insulator due to Proximity to a Topological Insulator, Phys. Rev. Lett. \textbf{125}, 017204 (2020). 
 
 \bibitem{Wang2023} Y. Wang, V. Lauter, O. Maximova, S. T. Konakanchi, P. Upadhyaya, J. Keum, H. Ambaye, J. Wang, M. Zhukovskyi, T. A. Orlova, B. A. Assaf, X. Liu, and L. P. Rokhinson, Exchange coupling in Bi$_2$Se$_3$/EuSe heterostructures and evidence of interfacial antiferromagnetic order formation, Phys. Rev. B \textbf{108}, 195308 (2023).

{\color{blue}
\bibitem{Wei2025} 
C.-C. Wei, X. Li, S. Hatt, X. Huai, J. Liu, B. Singh, K.-M. Kim, R. M. Fernandes, P. Cardon, L. Zhao, T. T. Tran, B. A. Frandsen, K. S. Burch, F. Liu, and H. Ji, La$_2$O$_3$Mn$_2$Se$_2$: A correlated insulating layered $d$-wave altermagnet, Phys. Rev. Materials \textbf{9}, 024402 (2025).
}

 
{\color{blue}
\bibitem{Zhu2026} Z. Zhu, R. Huang, X. Chen, Z. Cui, X. Duan, J. Zhang, I. \v{Z}uti\'{c}, and T. Zhou, Altermagnetic proximity effect, Phys. Rev. Lett. \textbf{136}, 186702 (2026).
}
  \bibitem{Krempasky2024} J. Krempask\'y, L. \v{S}mejkal, S. W. D'Souza, M. Hajlaoui, G. Springholz, K. Uhl\'i\v{r}ov\'a, F. Alarab, P. C. Constantinou, V. N. Strocov, D. Usanov, W. R. Pudelko, R. Gonzalez-Hernandez, A. B. Hellenes, Z. Jansa, H. Reichlov\'a, Z. \v{S}ob\'an, R. D. Gonzalez Betancourt, P. Wadley, J. Sinova, and T. Jungwirth, Altermagnetic lifting of Kramers spin degeneracy, Nature \textbf{626}, 517 (2024). 
  
  \bibitem{Reimers2024} S. Reimers, L. Odenbreit, L. \v{S}mejkal, V. N. Strocov, P. Constantinou, A. B. Hellenes, R. Jaeschke Ubiergo, W. H. Campos, V. K. Bharadwaj, A. Chakraborty, T. Denneulin, W. Shi, R. E. Dunin-Borkowski, S. Das, M. Kl\"aui, J. Sinova, and M. Jourdan, Direct observation of altermagnetic band splitting in CrSb thin films, Nat. Commun. \textbf{15}, 2116 (2024). 
  
  \bibitem{Regmi2025} R. B. Regmi, H. Bhandari, B. Thapa, Y. Hao, N. Sharma, J. McKenzie, X. Chen, A. Nayak, M. El Gazzah, B. G. M\'arkus, L. Forr\'o, X. Liu, H. Cao, J. F. Mitchell, I. I. Mazin, and N. J. Ghimire, Altermagnetism in the layered intercalated transition metal dichalcogenide CoNb$_4$Se$_8$, Nat. Commun. \textbf{16}, 4399 (2025).




\end{thebibliography}
\end{document}